# Quasi-flat bands in waveguide arrays

A.I. Maimistov[1,2]

[1]Department of Solid State Physics and Nanostructures, National Research Nuclear University MEPhI, Moscow, 115409, Russia
[2]Department of Physics and Technology of Nanostructures, Moscow Institute for Physics and Technology, Dolgoprudny, Moscow region, 141700 Russia.

aimaimistov@gmail.com

**Abstract**. Coupled electromagnetic waves propagating in a waveguide array are discussed. This waveguide array can be considered as one dimensional photon crystal composed from the unit cell containing three different waveguides, namely, two positive refraction index waveguides and one waveguide with negative refraction index. Spectrum of the linear modes of electromagnetic waves in this device contains the quasi-flat band.

## 1. Introduction

In the last twenty – twenty-five years new materials for optics have been fabricated. The best known example for these materials is the photon crystal or the photonic band-gap material [1] and metamatrerials with negative refractive index. The use of these materials allows investigating the different physics phenomena by optical simulation methods. For example, the optical simulation of Majorana physics was considered in [2] via two one-dimensional photonic crystals. In [3] the photonic realization of the quasi-particles carrying fractional statistics that interpolate between bosons and fermions was proposed. There are series of works devoted to photonic crystal waveguide analogy of the graphene [4, 5] and to the realization of Dirac point with double cones in optics [6-8]. Review of the optical analogies using photonic structures is presented in [9].

In the case of 2D photonic crystals under some conditions there is energy band that are pinned to Dirac points [10-14]. This band can be made very flat by controlling the parameters of the photonic crystals. It should be noted that flat band (or flat zone) is formed if the photon crystal unit cell contains three different "atoms". In the case under consideration "atom" is waveguide. Here waveguide array as 1D photonic crystal will be considered.

## 2. Alternating optical waveguide arrays

Two closely located waveguides can be coupled due to the tunneling of light from one waveguide to the other. This phenomenon forms the basis for device which is named coupler. A coupler using waveguides fabricated from materials with a positive refractive index (PRI) preserves the direction of light propagation, and for this reason it is named a directional coupler. If one of the waveguides of the coupler is fabricated from a material with a negative refractive index, then a radiation entering one waveguide leaves the coupler through the other waveguide at the same end but in the opposite direction [15]. For this reason this device was called the oppositely directional coupler.

Let us consider the waveguide array where waveguide having number $n$ is characterized by positive refractive index, whereas the nearest neighbouring waveguides with numbers $n-1$ and $n+1$ possess negative refractive index (NRI). If the electromagnetic radiation is localized in each waveguide the coupled wave theory can be used. The configuration of these alternating waveguides will be remarked as alternating optical waveguide arrays (AOWA). Dispersion relation for a linear wave in AOWA consists of two branch separated by a forbidden zone (energy gap). It is important to emphasize that the unit cell of AOWA contains two different waveguides, figure 1. In this figure $A_n$ ($B_n$) is slowly varying complex envelope of electric field in PRI-waveguide (NRI-waveguide) possessed by the *n*th unit cell.

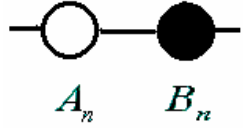

**Figure 1.** Unit cell of alternating optical waveguide arrays

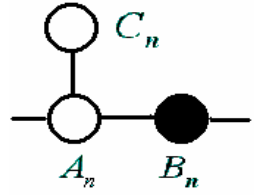

**Figure 2.** Unit cell of modified alternating optical waveguide arrays.

Let us suppose that waveguide array is composed from the unit cell containing three waveguides: two PRI-waveguides and one NRI-waveguide (figure 2). Here $A_n$ and $C_n$ are slowly varying complex envelopes of electric field in PRI-waveguides. The $B_n$ is slowly varying complex envelopes of electric field in NRI-waveguide. System of the normalized equations describing electromagnetic wave propagation in the this linear structure without taking into account the group velocity dispersion reads

$$i(A_{n,t} + A_{n,x}) + B_{n-1} + B_n + \beta C_n = 0, \qquad (1)$$

$$i(B_{n,t} - B_{n,x}) + A_n + A_{n+1} = 0, \qquad (2)$$

$$i(C_{n,t} + C_{n,x}) + \beta A_n = 0, \qquad (3)$$

It is suggested that the group velocities and the propagation constants are equal for all channels. Coordinate *x* is ration of the distance along waveguide to coupling distance between PRI-waveguide and NRI-waveguide. The coefficient $\beta$ is ration of coupling constant between neighbor PRI-waveguides to coupling constant between the PRI-waveguide and NRI-waveguide.

Let us suppose that waveguide array is rolled up to ring. It results in the following boundary conditions $A_{N+1} = A_1$, $B_{N+1} = B_1$ and $C_{N+1} = C_1$, where *N* is number of the ring cells. If the following ansatz

$$A_n(x,t) = a_n e^{-i\omega t + ikx}, \quad B_n(x,t) = b_n e^{-i\omega t + ikx}, \quad C_n(x,t) = c_n e^{-i\omega t + ikx},$$

is used, then the system of equations (1)-(3) reduces to equations

$$(\omega - k)a_n + b_{n-1} + b_n + \beta c_n = 0, \quad (\omega + k)b_n + a_n + a_{n+1} = 0, \quad (\omega - k)c_n + \beta a_n = 0. \quad (4)$$

This system of algebraic equations can be transformed to diagonal form by using the discrete Fourier transformation.

$$a_n = \sum_{s=-M}^{s=M} \tilde{a}_s \exp\{2\pi i s n / N\}, \quad b_n = \sum_{s=-M}^{s=M} \tilde{b}_s \exp\{2\pi i s n / N\}, \quad c_n = \sum_{s=-M}^{s=M} \tilde{c}_s \exp\{2\pi i s n / N\},$$

where $M = N/2$ or $M = (N-1)/2$ is integer number. Substitution of this expression into (4) results in the following system of equations

$$(\omega - k)\tilde{a}_s + z\tilde{b}_s + \beta \tilde{c}_s = 0, \quad (\omega + k)\tilde{b}_s + z^*\tilde{a}_s = 0, \quad (\omega - k)\tilde{c}_s + \beta \tilde{a}_s = 0, \quad (5)$$

where $z = 1 + \exp(-2\pi i s / N)$. The system of equations (5) has the nonzero solution only if the corresponding determinant is equal to zero. That leads to implicit form of the dispersion relation

$$(\omega - k)[(\omega^2 - k^2) - |z|^2] - \beta^2(\omega + k) = 0. \quad (6)$$

At $k = 0$ from (6) tree values of $\omega_s(0)$ follow: $\omega_s^{(1,2)}(0) = \pm(\beta^2 + |z|^2)^{1/2}$ and $\omega_s^{(0)}(0) = 0$. Hence three branches of the dispersion relation are present at each value of $s = 0, \pm 1, \pm 2, ..., \pm M$.

Figures 3 and 4 show several examples of the dispersion curves at $s = 0$.

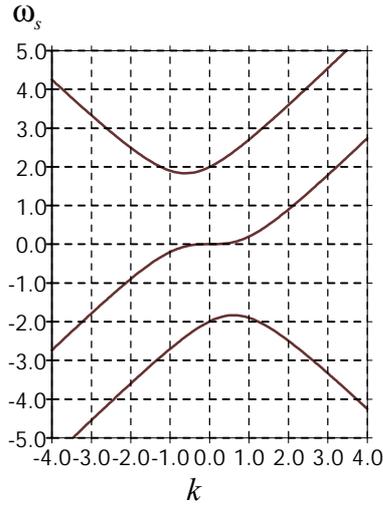 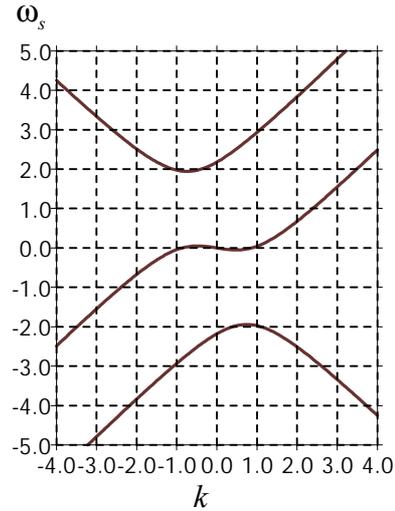

**Figure 3.** Dispersion curves at $|z|^2 = 2.0$ $\beta^2 = 1,98$

**Figure 4** Dispersion curves at $|z|^2 = 2.0$ $\beta^2 = 2,75$

If the coupling with third waveguide (i.e., in where the electric field is denoted as $C_n$) is equal to zero, then the dispersion equation (6) leads to famous result [16] for modes with gap

$$\omega_s^2(k) = |z|^2 + k^2,$$

and gapless mode $\omega_s(k) = k$.

Near point ($k = 0, \omega = 0$) the dispersion equation (6) can be solved approximately. This leads to dispersion relation for gapless mode

$$\omega_s(k) \approx \frac{|z|^2 - \beta^2}{|z|^2 + \beta^2} k .\qquad(7)$$

From equation (7) it follows that wave associated with gapless mode is akin to light wave in vacuum. However, the velocity of this wave can be very nearly equal to zero.

It is interesting to observe that in the case of infinite waveguide array (i.e., $N \to \infty$), at $k = 0$ dispersion relations are presented by following expressions

$$\omega^{(\pm)}(0,q) = \pm(\beta^2 + 4\cos^2 ql/2)^{1/2} \quad \omega^{(0)}(0,q) = 0 ,$$

where $q$ is the transverse wave number, $l$ is the lattice constant. It follows from this expressions that spectrum of the gapless mode belongs to flat band of 1D photon crystal. Because the true dispersion relation for gapless mode in not plane, we can say about quasi-flat band only.

## 3. Acknowledge

I would like to thank Dr. A.S. Desyatnikov, Dr. E.V. Kazantseva and A.O. Karavaeva for interesting and helpful discussions. The research was supported by Russian Scientists Found (project 14-22-00098).

**References**
[1] Yablonovitch E 1994 *J.Mod.Opt.* **41** 173
[2] Rodri'guez-Lara B M and Moya-Cessa H M 2014 *Phys.Rev.* **A89** 015803 [4 pages]
[3] Longhi S and Valle G D 2012 *Opt.Lett* **37** 2160
[4] Benisty H. 2009 *Phys.Rev.* **B79** 155409
[5] Mihalache I and Dragoman D 2011 *J.Opt.Soc.Amer.* **B28** 1746
[6] Sakoda K 2006 *J.Opt.Soc.Amer.* **B29** 2770
[7] Wang Li-Gang, Wang Zhi-Guo, Zhang Jun-Xiang and Zhu Shi-Yao 2009 *Optics Letters* **34** 1510
[8] Shen Ming, Ruan Lin-Xu, and Chen Xi 2010 *Optics Express* **18** 12779
[9] Longhi S 2009 *Laser & Photonics Rev.* **3** 243
[10] Apaja V, Hyrka"s M and Manninen M 2010 *Phys.Rev.* **A82** 041402(R)
[11] Jukic D, Buljan H, Lee D-H, Joannopoulos J D and Soljacic M 2012 *Optics Letters* **37** 5262
[12] Chui S T, Liu Shiyang and Lin Zhifang 2013 *Phys.Rev.* **E88** 031201(R)
[13] Chern Gia-Wei, Chien Chih-Chun, and Di Ventra M 2014 *Phys.Rev.* **A90** 013609
[14] Leykam D, Bahat-Treidel O and Desyatnikov A S 2012 *Phys.Rev.* **A86** 031805(R)
[15] Litchinitser N M, Gabitov I R and Maimistov A I 2007 *Phys. Rev. Lett.* **99** 113902
[16] Kazantseva E V, Maimistov A I and Ozhenko S S 2009 *Phys.Rev.* **A80** 43833 (7 pages)